# Two-Photon Spatial Multiplexing: Image Encoding in Gaussian Noncircularity of Light

Wanting Hou[1,2†], Zhiyuan Ye[1*†], Hong-Chao Liu[3], Jun Xiong[2*], Kaige Wang[2*]

[1]Department of Physics, Faculty of Arts and Sciences, Beijing Normal University, Zhuhai 519087, China.

[2]School of Physics and Astronomy, Applied Optics Beijing Area Major Laboratory, Beijing Normal University, Beijing 100875, China.

[3]Institute of Applied Physics and Materials Engineering, University of Macau, Avenida da Universidade, Taipa, Macau S.A.R., China.

*Corresponding authors. Email: yezy@bnu.edu.cn (Z.Y.); junxiong@bnu.edu.cn (J.X.); wangkg@bnu.edu.cn (K.W.).

†These authors contributed equally to this work.

## Abstract

The search for new degrees of freedom (DoF) of light is central to advancing optical information encoding and multiplexing. Gaussian statistics are ubiquitous in nature, and here we unlock an unexpected statistical DoF in the joint noncircularity of Gaussian light. For a pair of Gaussian light fields, the covariance- and pseudocovariance-dominated two types of correlations dictate the propagation rules and the image-encoding responses, mapping optical information into coordinate-difference and coordinate-sum momentum spaces, respectively. A hybrid random-phase grating array enables the experimental preparation of noncircular Gaussian light with programmable coexistence, yielding X-shaped intensity correlations while subtly reducing the degree of second-order coherence. Two independently programmed images coexist in the same beams at identical wavelength and polarization, yet are selectively decoded through two-photon correlation geometry without physical filters. Remarkably, both images remain recoverable even when structured stray light is as intense as the signal. These results unlock a hidden statistical dimension of optical information encoding, transforming Gaussian fluctuations into a dual-channel information carrier and joint noncircularity into a programmable optical DoF.

## INTRODUCTION

Gaussian statistics represents one of the most fundamental and universal statistical phenomena in nature and is widely encountered in optical systems (1). When many fluctuating contributions combine, the resulting optical field can often be described by a complex Gaussian random process. This description underlies much of the theory of thermal light, optical speckle, and partially coherent fields. Its practical power comes from a substantial reduction in statistical complexity: once the mean and all second-order field moments are specified, Gaussian moment factorization determines the higher-order moments (2). For zero-mean, jointly circular complex Gaussian fields, a prominent example is the connection between field coherence and intensity correlations in Hanbury Brown–Twiss (HBT) interferometry (3). This relation makes intensity fluctuations a route to spatial and temporal coherence without directly measuring the optical phase and also serves as a theoretical cornerstone for a range of phenomena and applications using thermal light (4–12), including ghost imaging, ghost diffraction, and subwavelength interference. Beyond coherence measurement and imaging applications, the degree of second-order coherence, $g^{(2)}$, also probes source and particle statistics (13–16) and light–matter dynamics (17–20). Together with higher-order correlations, such measurements can test Gaussian moment factorization and reveal non-Gaussian fluctuations (21–23).

Gaussianity, however, does not require circularity. A complex Gaussian field is completely characterized by its mean and the covariance $\Gamma = \langle E E^{\dagger} \rangle$ as well as the pseudocovariance $C = \langle E E^{T} \rangle$ of its fluctuations, which correlate conjugated and unconjugated amplitudes, respectively (2). Although this distinction is mathematically established, joint noncircularity remains little explored as a controllable resource in optics. Early optical studies examined noncircular Gaussian speckle generated by rough-surface scattering and its consequences for intensity contrast and phase statistics (24, 25). In the context of quantum and classical ghost imaging, Erkmen and Shapiro developed phase-sensitive coherence theory and a unified Gaussian-state framework (26), incorporating both conjugated and unconjugated cross-correlations and distinguishing propagation effects from signatures that require nonclassical light. Our recent work used random holography to generate conjugate-wavefront thermal-field pairs with programmable spatial correlations using linear optics (27, 28). These pairs display Einstein–Podolsky–Rosen (EPR)-like correlation geometries familiar from entangled photon pairs (29–32), while remaining classical and an inevitable correlation background. Each beam retains circular Gaussian statistics, but their conjugate random wavefronts create a nonzero cross pseudocovariance and hence a jointly noncircular Gaussian distribution. By contrast, conventional thermal light exhibits a nonzero covariance but a vanishing pseudocovariance. This distinction raises an interesting question whether the covariance and the pseudocovariance can coexist in noncircular Gaussian light.

Here we experimentally realize and control jointly noncircular Gaussian thermal light in which both the covariance and the pseudocovariance coexist. A hybrid random-phase grating independently programs the two statistical components while preserving circular Gaussian statistics in each beam. Under Fourier propagation, the covariance and pseudocovariance evolve into coordinate-difference and coordinate-sum spatial correlations, respectively, whose coexistence gives rise to the observed X-shaped intensity-correlation pattern. Dilution of the two components' normalized correlation amplitudes allows $g^{(2)}$ to fall below either isolated-channel value. When this jointly noncircular Gaussian light illuminates an object, the object's intensity-transmission structure and phase-dependent response are encoded in coordinate-difference and coordinate-sum correlations, respectively. A phase-shifted double-slit experiment illustrates this dual response. Their natural complementarity in spatial coordinates provides two separately addressable image-encoding channels. We demonstrate two-photon spatial multiplexing by encoding two distinct images at the same wavelength and polarization and selectively retrieving them through opposite- and same-direction joint detection, respectively, without requiring physical filters for channel selection. Both encoded images remain accessible in structured stray light. These results establish joint Gaussian noncircularity as a controllable statistical optical degree of freedom (DoF) for image encoding, complementing recent approaches based on polarization statistics (33) and skyrmion topology (34, 35).

## RESULTS

### Joint Gaussian statistics and complementary correlations

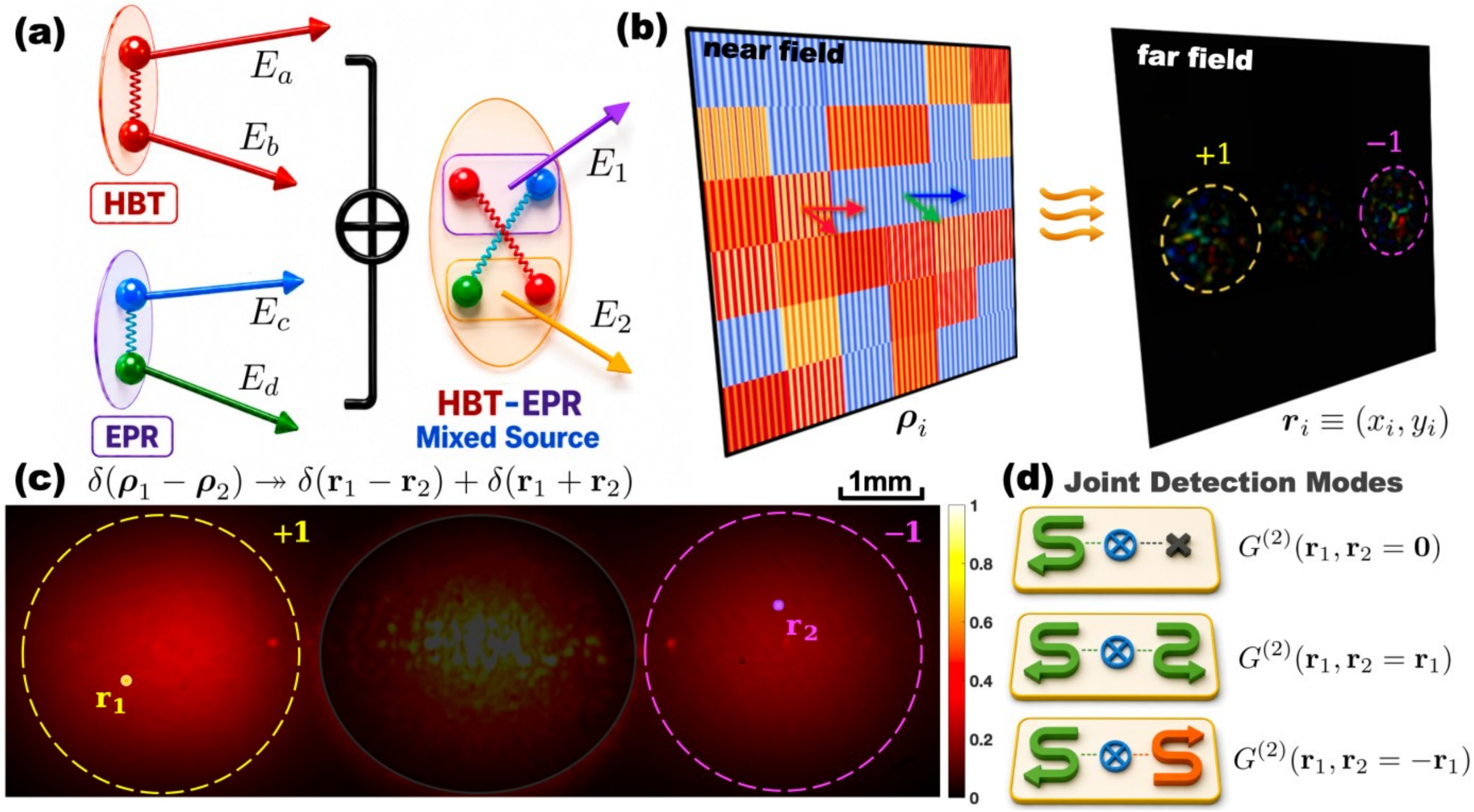

**Fig. 1. Generation of a mixed HBT–EPR-like source.** (a) Conceptual diagram. (b) Design of the hybrid random-phase grating array. (c) Ensemble-averaged intensity profile of the mixed source in the far-field Fourier plane over 5000 realizations. (d) Intensity correlation measurements of $G^{(2)}(r_1, r_2)$ performed between the ±1st-order diffracted beams in (c), for three joint spatial scanning modes.

As shown in Fig. 1(a), we prepare jointly noncircular Gaussian thermal light by combining two types of correlated field pair. In the conventional HBT-correlated pair $\{E_a, E_b\}$, the two fields share the same random wavefront. In the holographic thermal pair $\{E_c, E_d\}$, generated by random holography, their random wavefronts are complex conjugates (27, 28). Overlapping the corresponding fields gives $E_1 = E_a + E_c$ and $E_2 = E_b + E_d$. The two source families are statistically independent and share the same carrier wavelength and polarization. Here the random fields denote slowly varying transverse envelopes, and ensemble averages, denoted by angle brackets, are taken over programmed spatial-wavefront realizations. The fields have zero mean, and each individual beam follows circular complex Gaussian statistics. Their joint distribution can nevertheless be noncircular, also called improper (2, 36). For any field, we use intensity units in which $I_i = |E_i|^2$, and the mean intensity of mixed beam j is $N_j = \langle I_j \rangle > 0$. Spatial arguments are temporarily omitted, so each field is understood to be evaluated at the selected point in its own beam.

For the two mixed fields, the covariance matrix has entries $\Gamma_{ij} = \langle E_i E_j^* \rangle$, and the pseudocovariance matrix has entries $C_{ij} = \langle E_i E_j \rangle$, with $i, j = 1{,}2$. The indices label the beams: the off-diagonal elements $\Gamma_{12}$ and $C_{12}$ correlate amplitudes in beam 1 and beam 2, with and without complex conjugation, respectively. The measured second-order intensity correlation is the ensemble average of the product of the two intensities. Since each intensity contains two field amplitudes, this measurement is a fourth-order field moment. For jointly improper Gaussian fields, moment factorization expresses it as the sum of three products of second-order field moments (21, 26):

$$\begin{aligned} G^{(2)} &\equiv \langle I_1 I_2 \rangle = \langle E_1 E_1^* E_2 E_2^* \rangle \\ &= \langle E_1 E_1^* \rangle \langle E_2 E_2^* \rangle + \langle E_1 E_2^* \rangle \langle E_1^* E_2 \rangle \\ &+ \langle E_1 E_2 \rangle \langle E_1^* E_2^* \rangle \\ &= N_1 N_2 + |\Gamma_{12}|^2 + |C_{12}|^2 . \end{aligned} \quad (1)$$

The first term, $N_1 N_2$, is the mean-intensity background. The remaining terms describe correlations of the intensity fluctuations. The identical-wavefront construction supplies $\Gamma_{12}$ through $\langle E_a E_b^* \rangle$, whereas the conjugate-wavefront construction supplies $C_{12}$ through $\langle E_c E_d \rangle$; statistical independence removes cross moments connecting the two source families. We therefore call these the HBT (covariance) channel and the EPR-like (pseudocovariance) channel. Their coexistence makes the HBT–EPR-like mixed source a jointly noncircular Gaussian thermal source.

For the isolated holographic pair, the image-bearing term $|\langle E_c E_d \rangle|^2$ has the same modulus-squared form as the coincidence distribution $|\Psi|^2$ of a spatially entangled two-photon state, where $\Psi$ is the joint probability amplitude. Both joint amplitudes propagate with a product of two field-propagation kernels, leading to closely similar EPR-like spatial correlation structures (26, 27). The classical holographic measurement additionally contains the mean-intensity background $\langle I_c \rangle \langle I_d \rangle$. In the mixed source, the covariance channel contributes its own image-bearing term as well.

The distinction is a property of the joint statistics. Circular statistics in each beam require $C_{11} = C_{22} = 0$, while joint circularity requires every entry of the pseudocovariance matrix to vanish. A nonzero $C_{12}$ can therefore coexist with circular statistics in both individual beams. Adding independent Gaussian fields preserves Gaussianity and adds their covariance and pseudocovariance matrices. The simulated distributions in figs. S1 and S2 illustrate how the mixed pair remains Gaussian while acquiring joint noncircularity.

To compare measurements with different mean intensities, we divide the intensity-product average by the product of the mean intensities. Defining the normalized field moments $\gamma = \frac{\Gamma_{12}}{\sqrt{N_1 N_2}}$ and $\chi = \frac{C_{12}}{\sqrt{N_1 N_2}}$, Eq. (1) gives

$$\begin{aligned} g_{12}^{(2)} &\equiv \frac{\langle I_1 I_2 \rangle}{\langle I_1 \rangle \langle I_2 \rangle} = \frac{G^{(2)}}{N_1 N_2} \\ &= 1 + \frac{|\Gamma_{12}|^2 + |C_{12}|^2}{N_1 N_2} = 1 + |\gamma|^2 + |\chi|^2 . \end{aligned} \quad (2)$$

Equation (2) generalizes the conventional HBT relation $g_{12}^{(2)} = 1 + |\gamma|^2$ to jointly noncircular case. The two positive terms above this baseline quantify the contributions of the covariance and pseudocovariance channels. Their spatial patterns can be programmed independently, while their combined magnitude determines the normalized degree of second-order coherence.

## Second-order coherence and complementary propagation

Although the additional positive term in Eq. (2) might suggest enhanced intensity correlations, mixing the two source families can instead lower the normalized degree of second-order coherence because normalization by their combined mean intensities dilutes each channel's contribution. Let $I_{H,j}$ and $I_{E,j}$ be the respective mean intensities contributed to beam j, so that $N_j = I_{H,j} + I_{E,j}$. The correlations of the isolated source families, normalized by their own mean intensities, are $\rho_H = \frac{\langle E_a E_b^* \rangle}{\sqrt{I_{H,1} I_{H,2}}}$ and $\rho_E = \frac{\langle E_c E_d \rangle}{\sqrt{I_{E,1} I_{E,2}}}$. If the conventional family supplies the same fraction $\eta = \frac{I_{H,j}}{N_j}$ of the total mean intensity in both beams, the mixed-source correlation becomes

$$g_M^{(2)} = 1 + \eta^2 A + (1-\eta)^2 B, A = |\rho_H|^2, B = |\rho_E|^2. \tag{3}$$

The subscripts H, E, and M denote measurements with the conventional pair alone, the holographic pair alone, and the mixed source, respectively. Thus $g_H^{(2)} = 1 + A$ and $g_E^{(2)} = 1 + B$ are the isolated-channel values, and A and B are their excesses above the uncorrelated baseline of 1. For the resolved Gaussian-field model, $0 \le A, B \le 1$. When both excesses are positive, an appropriate intensity allocation lowers the mixed value below both isolated values. The minimum occurs at $\eta_* = B/(A+B)$, giving

$$1 < g_M^{(2)}(\eta_*) = 1 + \frac{AB}{A+B} < \min\{g_H^{(2)}, g_E^{(2)}\}. \tag{4}$$

The exact interval of fractions satisfying this strict inequality, for unequal isolated-channel excess correlations, is derived in Supplementary Section S2B. The reduction below both channels requires an appropriate intensity allocation; it does not occur for every mixing fraction. The contribution of each correlation is weighted by the square of its intensity fraction, so the normalized peak can weaken while both independently addressable image structures remain present.

At a common observation point where both constituent correlation responses reach their ideal unit values, $A = B = 1$, and the central peak obeys

$$g_M^{(2)}(0,0) = 1 + \eta^2 + (1-\eta)^2, \frac{3}{2} \le g_M^{(2)}(0,0) \le 2. \tag{5}$$

The minimum 1.5 occurs at equal intensities. This bound applies only at that common peak for ideal constituent pairs. It does not constrain arbitrary positions or imperfect source correlations. Adding two conventional pairs, or two holographic pairs, can instead retain a peak of 2 when their surviving field correlations are aligned (fig. S3). Thus correlation strength alone does not specify the number or geometry of the image channels.

The two channels also evolve differently during propagation. Let $h_j(r_j, \rho_j)$ describe how a field at source-plane position $\rho_j$ contributes to the field at detector-plane position $r_j$ in beam j. For this spatial calculation, $\Gamma(\rho_1, \rho_2)$ and $C(\rho_1, \rho_2)$ denote the cross moments $\Gamma_{12}$ and $C_{12}$ evaluated at two source-plane points, and primes denote the corresponding moments in the detector plane. Linear propagation gives

$$\begin{aligned} \Gamma'(r_1,r_2) &= \iint \Gamma(\rho_1,\rho_2) h_1(r_1,\rho_1) h_2^*(r_2,\rho_2)\, d\rho_1 d\rho_2, \\ C'(r_1,r_2) &= \iint C(\rho_1,\rho_2) h_1(r_1,\rho_1) h_2(r_2,\rho_2)\, d\rho_1 d\rho_2. \end{aligned} \tag{6}$$

In the covariance channel the second propagation kernel is complex conjugated; in the pseudocovariance channel it is not. This difference determines how the two spatial-correlation patterns evolve. For source-plane correlations localized near $\rho_1 = \rho_2$, matched Fourier propagation maps $\Gamma'$ onto a coordinate-difference response and $C'$ onto a coordinate-sum response. The corresponding Fourier-plane coordinate sums and differences are proportional to the transverse momentum sums and differences. The coordinate-difference response forms a ridge along equal detector coordinates, whereas the coordinate-sum response forms a ridge along opposite coordinates. Their superposition produces the far-field X-shaped correlation. Finite source apertures set the ridge widths, as derived in Supplementary Sections S1B to S1D. When the same target $T$ is inserted, the two channels sample the Fourier intensities of $|T|^2$ and $T^2$, respectively (Supplementary Section S3); the latter retains twice the object phase.

### Generation of the mixed Gaussian source

Experimentally, the mixed source is synthesized with a composite random phase-grating array [Fig. 1(b)]. Each elementary sinusoidal phase grating can be written as

$$t(x) = \exp\left[i\phi_0 + i m \sin\left(2\pi x/d + \phi_1\right)\right], \tag{7}$$

where $m$ is the phase-modulation depth, $d$ is the grating period, $\phi_0$ and $\phi_1$ are two types of random initial phases ranging from 0 to $2\pi$. Using the Jacobi-Anger identity, one obtains the complex amplitude of the $q$th-order expansion $A_q = J_q(m)\exp[i(\phi_0 + q\phi_1)]$, where $J_q$ is the Bessel function of order $q$. In particular, the $\pm 1$st diffraction orders carry opposite signs of $\phi_1$ while $\phi_0$ is independent of the diffraction order. As shown in Fig. 1(b), the two random phase encoding patterns are spatially randomly mixed: in the red regions $\phi_0$ is random while $\phi_1$ is fixed at 0, producing a pair of random subwaves with a constant wavefront difference; in the blue regions $\phi_1$ is random while $\phi_0$ is fixed, producing a pair of random subwaves with a constant wavefront sum. By combining many such randomly phased grating elements and selecting appropriate $\pm 1$st diffraction-order pairs (Fig. 1(c)), under coherent illumination, this grating array provides the identical-wavefront and conjugate-wavefront building blocks required for the jointly noncircular Gaussian thermal source. Such a grating array can be fabricated as a mask using techniques such as photolithography, or it can be flexibly implemented using two cascaded spatial light modulators (see Materials and Methods for details).

### Intensity correlation and double-slit interference

Before demonstrating multiplexing, we characterize the mixed source in the far-field observation plane, an optical Fourier-transform plane formed by a lens of focal length $f$. Two-point correlations are evaluated between regions around the $\pm 1$ diffraction orders in Fig. 1(c). Sampling the central rows yields the X-shaped $g^{(2)}(x_1, x_2)$ in Fig. 2(a), using 5000 realizations. The measured intersection peak is $1.29 \pm 0.04$, below the ideal equal-weight value of 1.5 in Eq. (5). The intrinsic reduction from 2 to 1.5 follows from the squared fractions in Eq. (3), whereas the further reduction reflects the measured isolated-channel excess correlations and experimental imperfections. The separately measured peaks are $1.77 \pm 0.04$ for the conventional channel and $1.54 \pm 0.04$ for the holographic channel (fig. S7). With the implemented 45:55 subgrating ratio and comparable diffraction efficiencies, Eq. (3) predicts $g_M^{(2)} \simeq 1.32$, consistent with the mixed-source measurement. Source preparation and balancing are described in Materials and Methods.

Now we show how the spatial information is encoded into the X-shaped correlation of the mixed source. Here we take the double slit $T$ as an example. Two-photon double-slit interference has been studied with entangled photon pairs (37–39) or with thermal light (4, 5, 40). As shown in Fig. 2(b), the double slit is placed at the source plane, and the joint correlation detection is performed in the far-field plane to yield the connected, or background-subtracted, correlation function ( $G_c^{(2)} \equiv G^{(2)} - \langle I_1 \rangle \langle I_2 \rangle$) which measures correlated intensity fluctuations and contains the interference-diffraction patterns of $T(\boldsymbol{\rho})$

$$\begin{aligned} G_c^{(2)}(x_1, x_2) &\propto A_H \left| \mathscr{F}\{|T(\rho)|^2\} \right|^2 + A_E \left| \mathscr{F}\{T^2(\rho)\} \right|^2 \\ &\propto A_E \operatorname{sinc}^2\left[\frac{\pi w}{\lambda f}(x_1 + x_2)\right] \cos^2\left[\frac{\pi s}{\lambda f}(x_1 + x_2) - \Delta\phi\right] \\ &+ A_H \operatorname{sinc}^2\left[\frac{\pi w}{\lambda f}(x_1 - x_2)\right] \cos^2\left[\frac{\pi s}{\lambda f}(x_1 - x_2)\right], \end{aligned} \tag{8}$$

Here $w$, $s$, and $\Delta\phi$ are the slit width, center-to-center spacing, and relative slit phase, while $A_E \approx A_H$ describe the nearly balanced contributions of the two source families. According to Eq. (8), the HBT (covariance) channel gives the squared magnitude of the Fourier transform of the intensity transmission, $|T|^2$, whereas the EPR-like (pseudocovariance) channel gives the squared magnitude of the Fourier transform of $T^2$. The latter depends on spatial variations in twice the target phase. The two patterns depend on the coordinate difference $x_1 - x_2$ and coordinate sum $x_1 + x_2$, respectively. Choosing how the two observation coordinates vary together therefore selects which fringe pattern is read.

The slit separation is much larger than the transverse coherence length of the mixed source, so the ensemble-averaged intensity shows no visible interference fringes. For $\Delta\phi = 0$, fixing one observation point at $x_2 = 0$ and varying the other recovers fringes in the connected correlation $G_c^{(2)}(x, 0)$ [Fig. 2(c)]. Sampling the two positions together, with either $x_1 = x_2$ or $x_1 = -x_2$, doubles the coordinate sum or difference in Eq. (8). The resulting correlation fringes have half the period of the fixed-coordinate scan, corresponding to the familiar $\frac{\lambda}{2}$ two-photon interference scale (38, 39, 4). When the object phase difference $\Delta\phi$ is nonzero, the two channels have the same fringe period but different initial phases. At $\Delta\phi = \frac{\pi}{2}$, their modulations are complementary. For equal effective channel weights, Eq. (8) predicts a $sinc^2$ envelope without fringe modulation when one observation point is fixed at $x_2 = 0$. The nearly balanced experiment in Fig. 2(d) shows suppression of this fixed-coordinate fringe modulation, while the selected joint scans retain the channel-specific patterns.

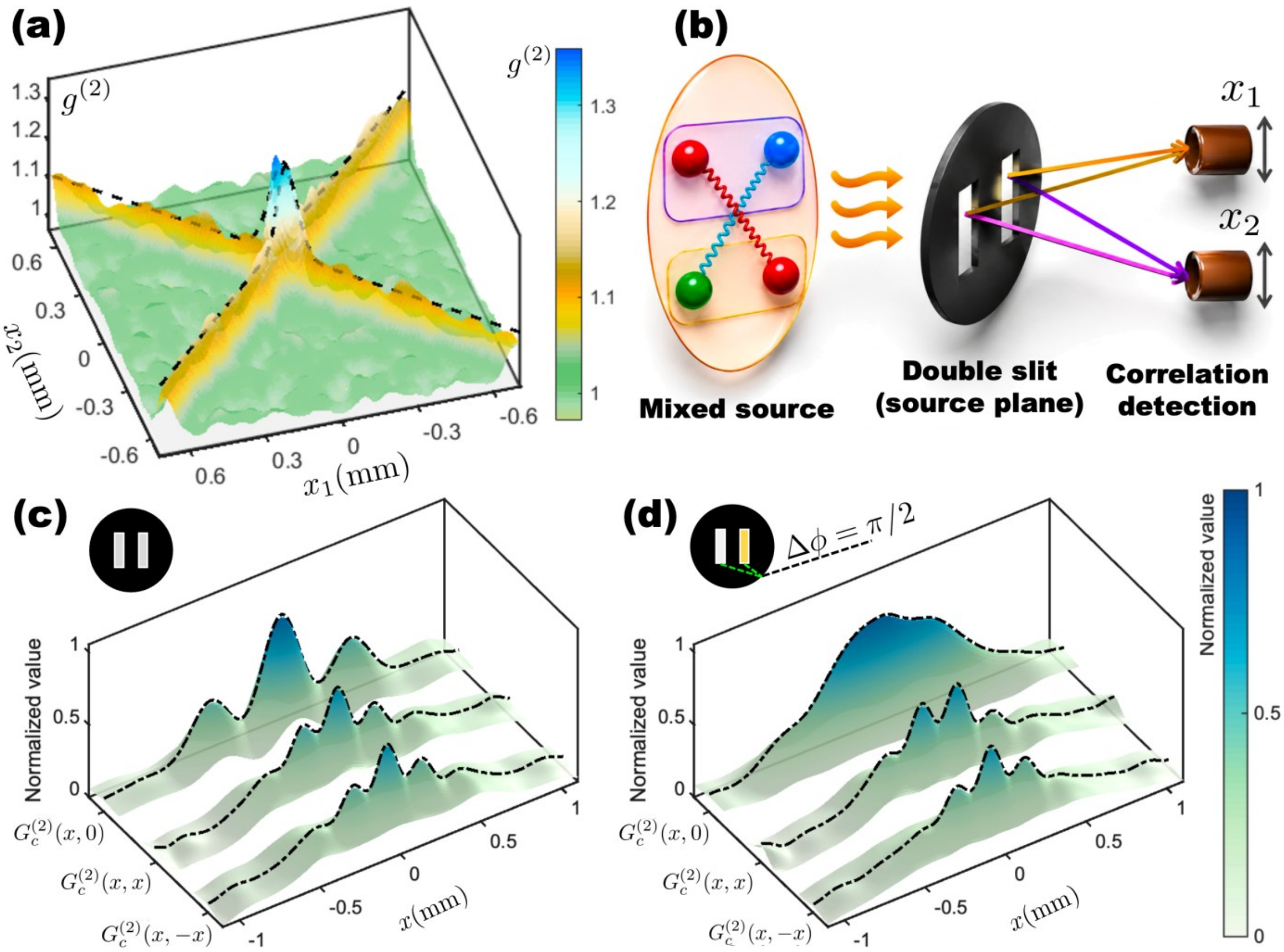

**Fig. 2. Correlation and double-slit interference measurements.** Experimental observations of (a) far-field normalized correlation function $g^{(2)}(x_1, x_2)$ (5000 realizations) and (b-d) double-slit interference with the mixed source (10 000 realizations). Each correlation curve in (c)(d) is displayed after subtracting the intensity background and normalizing to [*0*,*1*].

The mixed source makes two different target-dependent responses accessible within the same field pair. For an amplitude-only object, $T^2 = |T|^2$, so the two responses can contain redundant information. For a complex-amplitude target, the two channels encode complementary information about its intensity transmission and phase-dependent response. Supplementary Section S5 and fig. S11 illustrate improved reconstruction in a numerical example using both constraints. As a result, the reduced correlation peak of the mixed source nevertheless allows the target's dual-information recovery.

### Two-photon spatial multiplexing and demultiplexing

To encode two independent images, we apply separate complex-amplitude modulations $T_i(\rho)$ to the four component fields through $E_i \to E_i T_i$, for $i = a, b, c, d$ [Fig. 3(a)]. In the double-slit experiment, the same target acts on both correlation channels. Here, each channel receives its own programmed modulation before the source components are spatially superposed (see Materials and Methods).

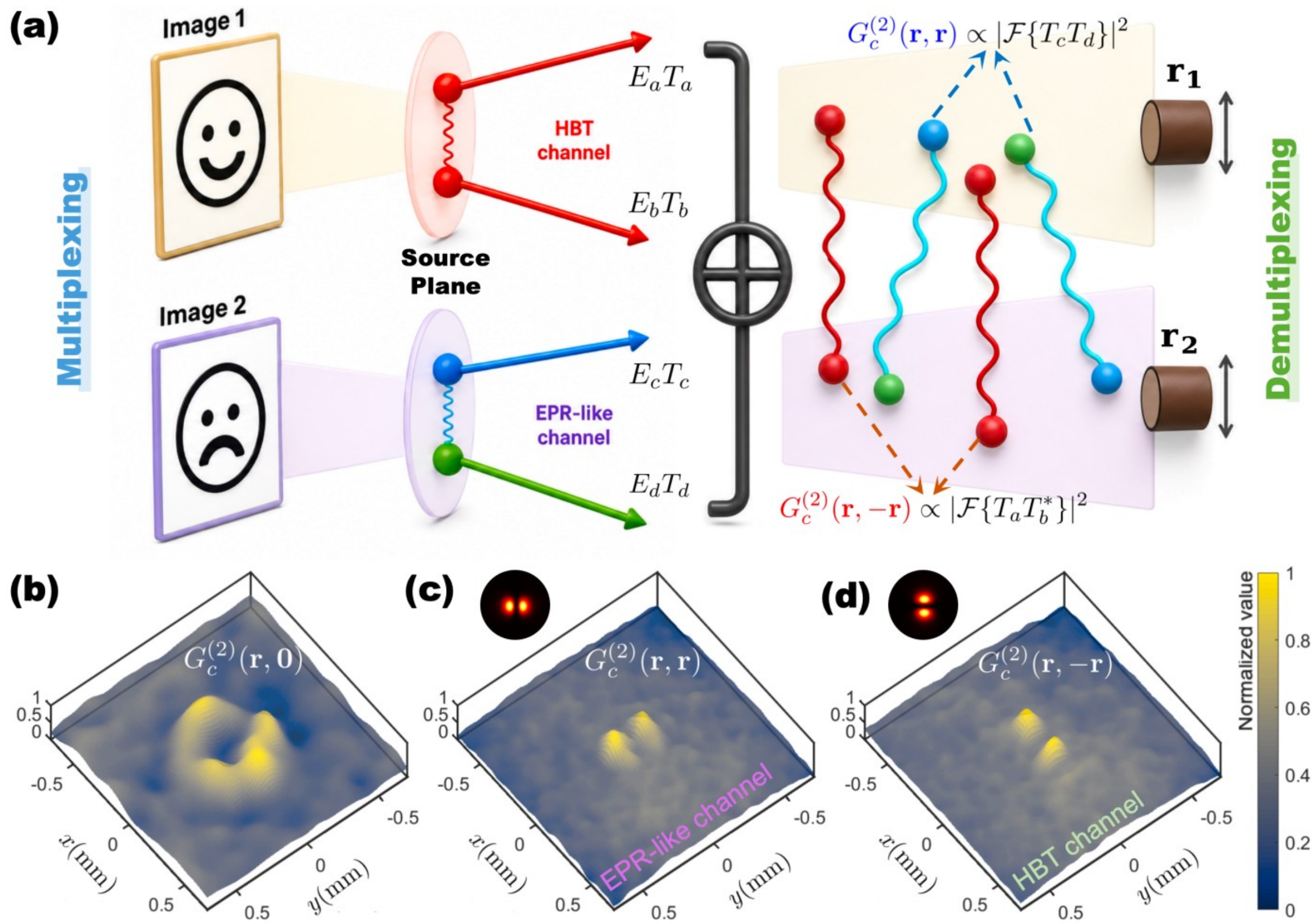

**Fig. 3. Two-photon spatial multiplexing inspired by cylindrical vector beams.** (a) Conceptual diagram. (b) With one detection coordinate fixed, the two encoded images, designed to reproduce orthogonal component patterns of a radially polarized beam, superpose into a ring. (c) Same-direction sampling selects the pseudocovariance-channel image. (d) Opposite-direction sampling selects the covariance-channel image. Both channels share the same wavelength and polarization, and no physical filter is used for channel selection. Panels (b) to (d) use the three joint-detection modes in Fig. 1(d), with 200,000 realizations for each correlation pattern.

After propagation, the connected correlation function in the far field can then be obtained

$$G_c^{(2)}(r_1,r_2) \propto \left|\tilde{T}_{ab}\left(\frac{r_1-r_2}{\lambda f}\right)\right|^2 + \left|\tilde{T}_{cd}\left(\frac{r_1+r_2}{\lambda f}\right)\right|^2, \tag{9}$$

Here $\tilde{T}_{ab} \propto \mathscr{F}\{T_a T_b^*\}$ and $\tilde{T}_{cd} \propto \mathscr{F}\{T_c T_d\}$. The covariance channel carries the product $T_a T_b^*$, whereas the pseudocovariance channel carries $T_c T_d$. Independent source-plane phase patterns therefore write distinct images into two statistical components of the same detected field pair, with no distinction in carrier wavelength or polarization. Their coordinate-difference and coordinate-sum responses remain separately addressable after the source families are superposed. A phase-retrieval algorithm (41) determines the phase modulation needed to reproduce each target image (implementation details are given in Materials and Methods).

The image design in Fig. 3(b) is inspired by cylindrical vector beams, whose orthogonal polarization components carry different spatial intensity patterns. We program the two correlation channels to reproduce the orthogonal component patterns of a radially polarized beam (42, 43). Fixing one detection coordinate and varying the other superposes the two correlation images, producing the ring-shaped distribution in Fig. 3(b). This provides a statistical analogue of the image superposition encountered in vector light and polarization multiplexing, while both encoded channels here occupy beams with the same fixed polarization.

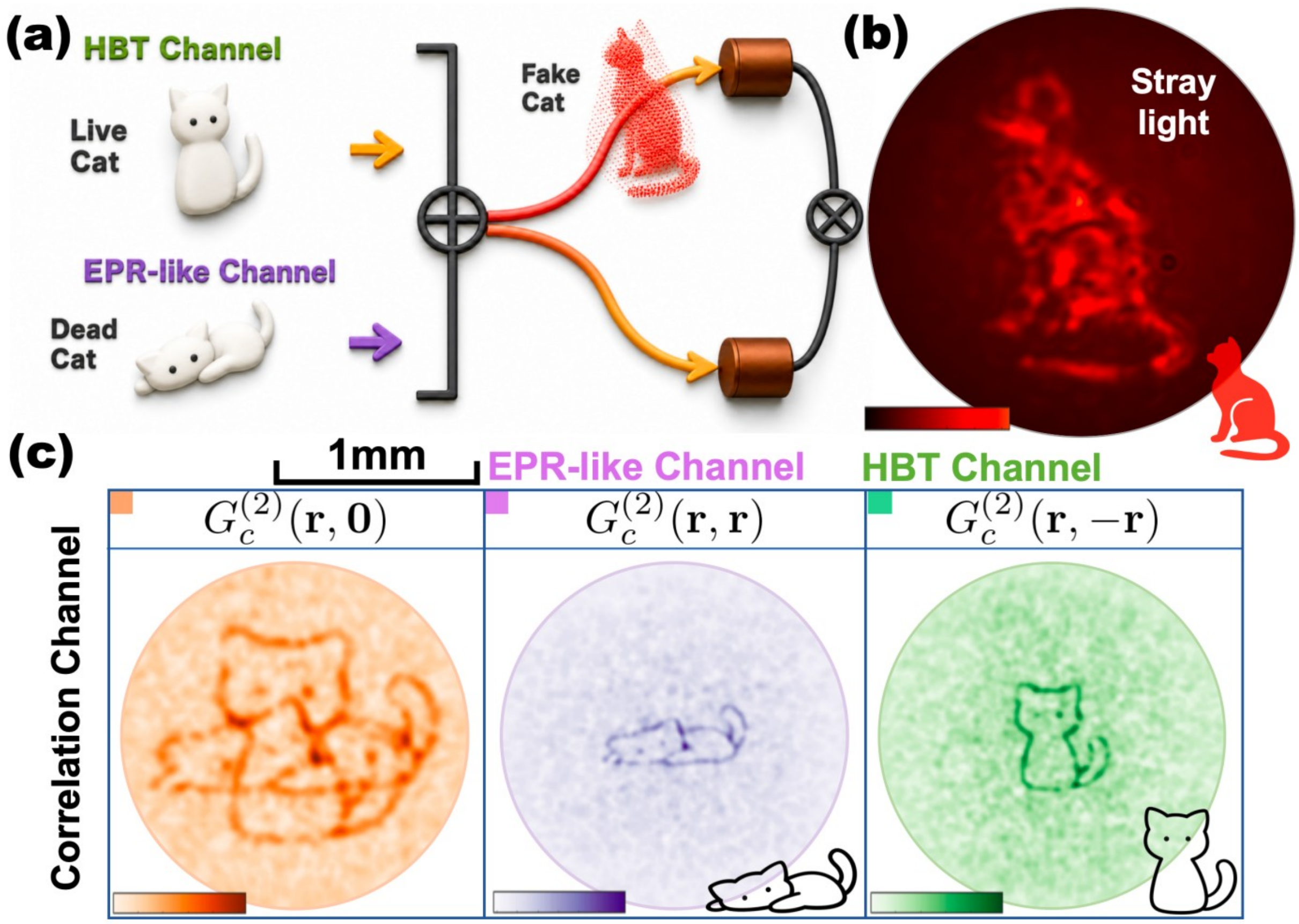

**Fig. 4. Correlation-channel recovery in the presence of stray light.** (a) Conceptual diagram: based on Fig. 3(a), a noisy stray light beam imprinted with a cat pattern is introduced into one beam. (b) Strong stray light obscures the direct ensemble-averaged intensity distribution in one arm. (c) The two-point correlation measurements recover the encoded images without a resolved stray-pattern artifact based on three types of joint-detection modes ($10^6$ realizations). All displayed images are normalized to [*0*,*1*].

The images are selected by joint detection geometry. Same-direction sampling, $r_1 = r_2$, varies the pseudocovariance response and selects the image encoded in the EPR-like channel [Fig. 3(c)], while the HBT-channel contribution becomes mere background and supplies a spatially uniform offset along this trajectory. Opposite-direction sampling, $r_1 = -r_2$, selects the covariance-channel image [Fig. 3(d)] and fixes the EPR-like contribution. The selected patterns have a transverse scale compressed by a factor of two because the sampled coordinate sum or difference doubles along either trajectory. Demultiplexing therefore requires no physical filter for channel selection, and the unselected channel is spatially constant along the ideal sampling trajectory rather than physically removed.

The proposed correlation-channel decoding also provides a natural discrimination mechanism against uncorrelated stray-light background. In the final demonstration case [Fig. 4(a)], the patterns of a live cat and a dead cat are assigned to the HBT and EPR-like channels, respectively, while another cat-shaped noisy stray light beam is introduced into one arm of the mixed source as shown in Fig. 4(b). The mean stray-light-to-signal intensity ratio is approximately unity (0 dB). The mean intensity reveals the added stray-light pattern, while the two encoded images are absent from the mean-intensity structure. The three joint detection modes recover the live-cat image, the dead-cat image, or their superposition in Fig. 4(c), using $10^6$ realizations. No resolved stray-pattern artifact is present in the recovered correlation images under these conditions. A statistically independent background principally increases the mean-intensity background and finite-sample noise rather than reproducing its structure in the connected correlation. The experiment therefore distinguishes information visible in the average intensity from information stored in correlated intensity fluctuations. Interestingly, the results in Fig. 4 can also be regarded as the multiplexing of three optical images: one encoded in the first-order intensity average (stray light), and the other two encoded in two complementary second-order intensity correlations. It is found that all three can be extracted without crosstalk.

## DISCUSSION

Our experiments turn joint Gaussian noncircularity into an addressable statistical DoF for encoding two independent images within one pair of optical fields. The covariance and pseudocovariance carry separately programmed information at the same wavelength and polarization. Phase-retrieval holograms (41) write the images, and the cylindrical-vector-beam-inspired demonstration (42, 43) shows how their superposition can be resolved through joint sampling. The established complex Gaussian theory supplies the statistical description (2, 26, 36). This work also reveals the non-negligible role of joint non-circularity in Gaussian correlations, which not only directly determines the degree of second-order coherence (a reduction in theory from 2 to 1.5) but also reshape how second-order correlations evolve in space. The noncircular HBT relation shows that a mismatch between $g^{(2)}$ and the first-order coherence function (44) does not necessarily imply non-Gaussianity: joint noncircularity must also be examined, because a nonzero pseudocovariance contributes to intensity correlations even within the Gaussian framework.

The present work extends optical multiplexing from field-level DoFs to the geometry of two-photon correlations. Recent representative single-channel image-hiding schemes place an image in either quantum or classical spatial correlations, making it inaccessible to direct intensity measurements (45–49). Our work demonstrates the controllable coexistence and dual-channel multiplexing of the complementary covariance and pseudocovariance channels. Demultiplexing is then achieved by choosing the joint detector-scanning trajectory, which exploits their complementary coordinate-difference and coordinate-sum structures in the joint far-field correlation space. Both channels originate from the same coherent source and thus share the same wavelength and polarization. Within this operating regime, the results establish a direct route from Gaussian noncircularity to optical image multiplexing: complementary field moments carry independent information, and correlation geometry provides the means to select it. The protocol can therefore be integrated with multiplexing schemes based on other optical DoFs.

The two-photon spatial multiplexing scheme here is realized using classical engineered pseudothermal light fields. The EPR-like branch could, in principle, be replaced by a spontaneous parametric down-conversion (SPDC) source of spatially entangled photon pairs, with correlations measured using single-photon detectors. More generally, the protocol requires two distinguishable and independently addressable spatial correlation channels, associated with the coordinate difference and coordinate sum, respectively. Spatial entanglement is therefore a possible resource for realizing the sum-coordinate channel (50), rather than a prerequisite for two-photon spatial multiplexing. This possibility also raises a further question: because SPDC source and holographic thermal light both encode images in coordinate-sum space, can their image contributions be selectively separated and distilled (51–53) when the two sources coexist? Addressing this question would require a distinguishing mechanism beyond the correlation geometry exploited here.

## MATERIALS AND METHODS

### Study design

The study tested whether two independently addressable spatial-correlation channels could coexist in a pair of classical pseudothermal fields with locally circular Gaussian statistics. We first characterized the mixed source and its double-slit response, then encoded and selectively recovered independent images, including in structured stray light. The experiment measured spatial intensity correlations and channel-dependent propagation and phase responses.

### Optical setup and double-slit preparation

All experimental observations are based on the optical layout in Fig. 5. A He-Ne laser beam (wavelength $\lambda = 632.8$ nm) is collimated and expanded (not shown, the beam diameter is ~2 mm), then passes through a polarizer and a half-wave plate before illuminating a transmissive spatial light modulator ($TSLM_1$, Daheng Optics, GCI-770104, pixel pitch: 12.5 μm). Combined with a subsequent 4f imaging system and an aperture placed in the Fourier plane (allowing only the +1st-order diffracted beam to pass), an arbitrary structured beam can be generated and then imaged onto the reflective SLM ($RSLM_1$, Holoeye, PLUTO-2-NIR-015, pixel pitch is 8 μm). In most experiments, only a holographic grating is loaded onto $TSLM_1$ with the diffraction angle of $0.71^{\circ}$ (the angle between 0th and +1st-order diffraction). In the experiments shown in Figs. 2b–d, holograms corresponding to two kinds of double slits are loaded, with a slit separation of $s = 425$ μm and a slit width of $w = 212.5$ μm in Supplementary Eq. (S48). All optical Fourier lenses in the setup have an identical focal length of $f = 30$ cm.

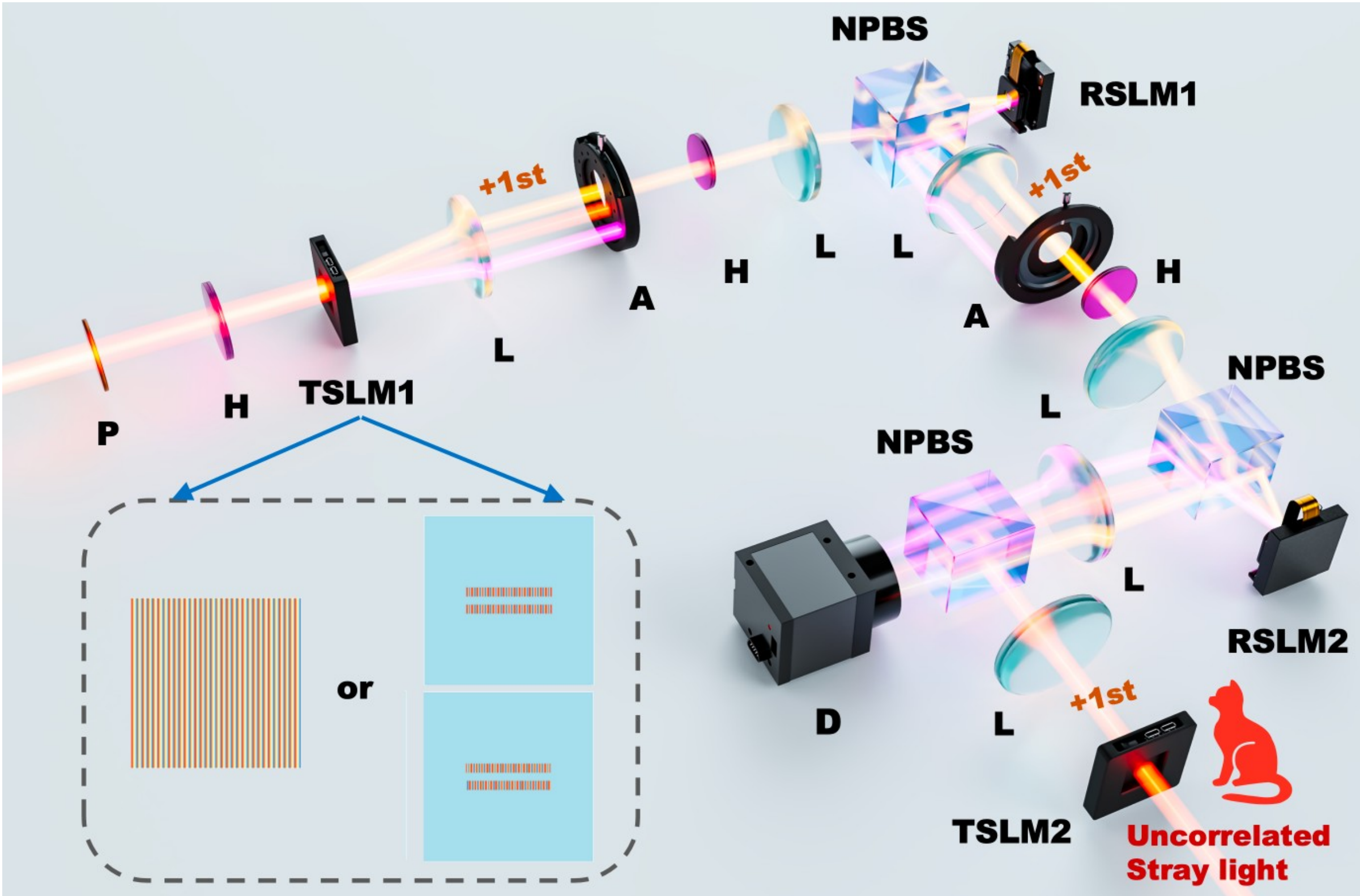

**Fig. 5. Experimental setup.** P: polarizer; H: half-wave plate; TSLM and RSLM: transmissive and reflective spatial light modulator; A: aperture diaphragm; L: Fourier lens (*f*=*30* cm); NPBS: non-polarizing beam splitter; D: CMOS-based image sensor.

### Mixed source generation with cascaded spatial light modulators

For more precise phase-only modulation and to facilitate the subsequent optical multiplexing experiments, two RSLMs are employed here to prepare the mixed HBT-EPR-like thermal source. The two RSLMs are configured through a 4f imaging system, and an aperture is used to allow only the +*1*st-order diffracted beam from $RSLM_1$ to pass. Figure 6 illustrates the procedure for a pair of hologram patterns loaded onto the two RSLMs. To optimize the diffraction efficiency, the two RSLMs employ different holographic encoding methods. It should be noted that, different from the procedure in Fig. S4, the two RSLMs are not used to separately generate HBT-type pseudo-thermal light or holographic thermal light. By adding the phase $\psi_1$ in Fig. 6 and the phase $\psi_0$ (corresponding to the blazed grating), scaling the resulting hologram to the operating phase range of $RSLM_1$, and loading it, a spatially partially random wavefront can be reproduced in the +*1*st-order diffracted beam. $RSLM_2$, on the other hand, is loaded with a sinusoidal phase hologram with a modulation depth chosen for efficient ±1-order diffraction. The phase-modulation depth is adjusted for each RSLM to optimize its diffraction efficiency. Most of the energy is allocated to the +1st-order diffracted beam for $RSLM_1$. In contrast, $RSLM_2$ must allow both ±*1*st-order diffracted beams to be measured simultaneously, so the energy allocated to the two diffraction orders should be strong and comparable. The random wavefront from $RSLM_1$ illuminates the deterministic grating corresponding to the plane-wave region on $RSLM_2$, thereby generating a spatially separated HBT-type pseudo-thermal light beam pair, while the plane-wave region from $RSLM_1$ illuminates the random grating region on $RSLM_2$, thereby producing a holographic thermal light pair. The grating periods of the two RSLMs are set to be identical, with both diffraction angles equal to $0.40^{\circ}$. Consequently, upon propagation, the two components mix and coherently superpose in the far field. The two cannot be distinguished by direct intensity measurements, and their difference becomes evident only through joint intensity correlation measurements. For the experiment of X-shaped correlation measurement in Fig. 2(a), the size of the sub-gratings in the random grating array is set to *15×15* pixels, i.e., occupying an area of *120* μm×*120* μm. In the other experimental observations, the sub-grating size is set to *10×10* pixels, i.e., occupying an area of *80* μm×*80* μm.

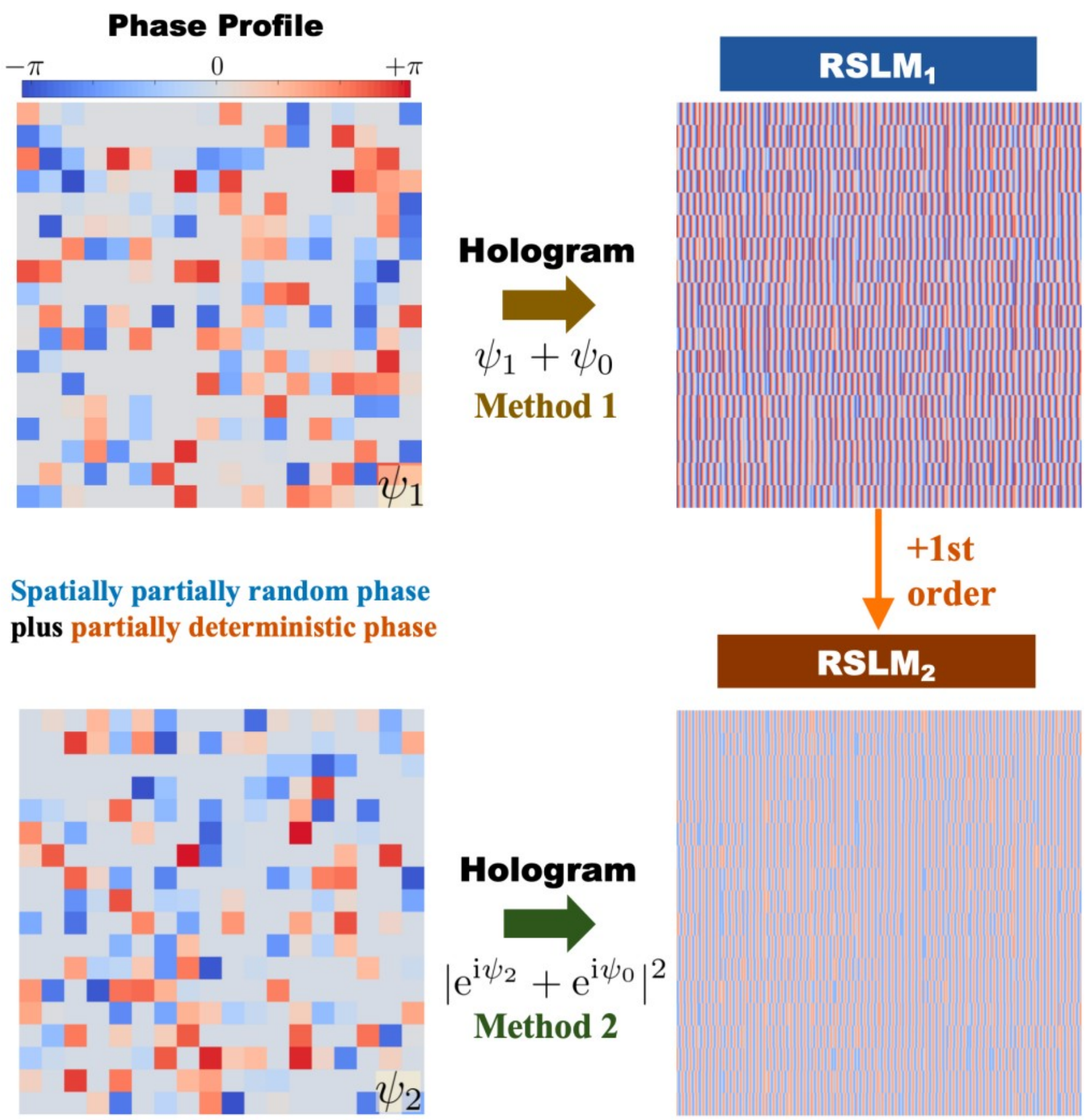

**Fig. 6. Generation method of the mixed HBT-EPR-like pseudo-thermal light field using two cascaded RSLMs.** Note that, for precise alignment, the hologram pattern loaded on $RSLM_2$ needs to be rotated by 180° (not shown here) to compensate for the image inversion introduced by the 4f imaging system.

### Intensity acquisition and correlation processing

A CMOS image sensor (Daheng Optics, MER-231-41U3M; pixel pitch, 5.86 μm) is placed in the Fourier plane of $RSLM_2$ to record both ±1 diffraction orders simultaneously. Their small diffraction angle allows both beams to fit on the same sensor. The two RSLMs synchronously display random holograms at 20 Hz, while the camera records intensity patterns at 10 Hz. Each exposure is much shorter than the hologram display interval. Intensity correlations are then calculated by averaging products of pixel intensities over the recorded realizations.

The full function $G^{(2)}(r_1, r_2)$ depends on four spatial coordinates because each position $r_j$ has two transverse components. Rather than reconstructing every position pair, we evaluate the three two-dimensional sampling trajectories in Fig. 1(d). The origin in each beam is its center in the ensemble-averaged intensity image. Fixed-coordinate sampling correlates each selected pixel in one beam with the central pixel of the other. Same-direction and opposite-direction sampling correlate pixel pairs at $r_2 = r_1$ and $r_2 = -r_1$, respectively. Thus the joint scans are implemented by selecting pixel pairs in the recorded camera images.

We use two representations of the intensity correlations. Dividing $G^{(2)} = \langle I_1 I_2 \rangle$ by the product of the mean intensities gives the dimensionless degree of second-order coherence, $g^{(2)} = \frac{\langle I_1 I_2 \rangle}{\langle I_1 \rangle \langle I_2 \rangle}$, shown in Fig. 2(a). Subtracting this mean-intensity background instead gives the connected correlation, $G_c^{(2)} = \langle I_1 I_2 \rangle - \langle I_1 \rangle \langle I_2 \rangle = \langle \delta I_1 \delta I_2 \rangle$, where $\delta I_j = I_j - \langle I_j \rangle$. This quantity isolates the spatial structure carried by correlated intensity fluctuations. For image display, the background-subtracted patterns are additionally rescaled to [0,1]; this display rescaling does not preserve the absolute value of $g^{(2)}$.

## Correlation calibration and channel balancing

The measured intersection peak in Fig. 2(a), $1.29 \pm 0.04$, is below the ideal equal-intensity value of 1.5 because the ideal calculation assumes perfectly correlated constituent pairs. The isolated HBT and EPR-like pairs measured in the same setup instead give peaks of $1.77 \pm 0.04$ and $1.54 \pm 0.04$, respectively (fig. S7). Possible sources of reduced measured bunching include detection noise, imperfect alignment of the RSLMs, and departures from unit magnification in the 4f imaging systems. The EPR-like component can also be affected by accumulated imaging errors and the incident wavefront: its correlation response contains twice the incident phase, whereas the common incident phase cancels from the conventional HBT response (27). The reported peak values are the means of five measurements, and the quoted uncertainties are their standard deviations.

To make the two recovered image signals comparable, we allocate 45% of the subgratings to the conventional family and 55% to the holographic family. If their diffraction efficiencies are comparable, this allocation corresponds approximately to $\eta = 0.45$. The measured isolated-channel excesses above the uncorrelated baseline $g^{(2)} = 1$ are 0.77 and 0.54. Using these measured excesses as effective inputs to Eq. (3), or Supplementary Eq. (S23), predicts an intersection peak of 1.32, consistent with $1.29 \pm 0.04$.

## Phase encoding for two-photon spatial multiplexing

Image multiplexing requires the modulation $T_i$ to be applied to each source component before the components are spatially superposed [Supplementary Eq. (S40)]. In the two-RSLM implementation, deterministic phase patterns replace the plane-wave regions shown in Fig. 6, as illustrated in Fig. 7. We calculate the phase pattern for each desired Fourier-plane image using the hybrid input–output/error-reduction (HIO/ER) phase-retrieval algorithm (41), iterated hundreds of times.

The optimized phase must be divided by two before it is loaded because each correlation channel contains a product of two field modulations. For the EPR-like channel, a phase $\phi_{cd}$ loaded on RSLM1 gives $T_c = T_d = \exp(i \phi_{cd})$, so the encoded product is $T_c T_d = \exp(i 2 \phi_{cd})$ [Supplementary Eq. (S43)]. For the HBT channel, a phase $\phi_{ab}$ loaded on RSLM2 gives $T_a = \exp(i \phi_{ab})$ and $T_b = \exp(-i \phi_{ab})$, so $T_a T_b^* = \exp(i 2 \phi_{ab})$ [Supplementary Eq. (S42)]. Loading half the optimized phase therefore reproduces the intended image in either channel. $RSLM_1$ carries the deterministic image phase for the EPR-like channel and the random-wavefront seed for the HBT component. $RSLM_2$ carries the deterministic image phase for the HBT channel and the random-grating seed for the EPR-like component.

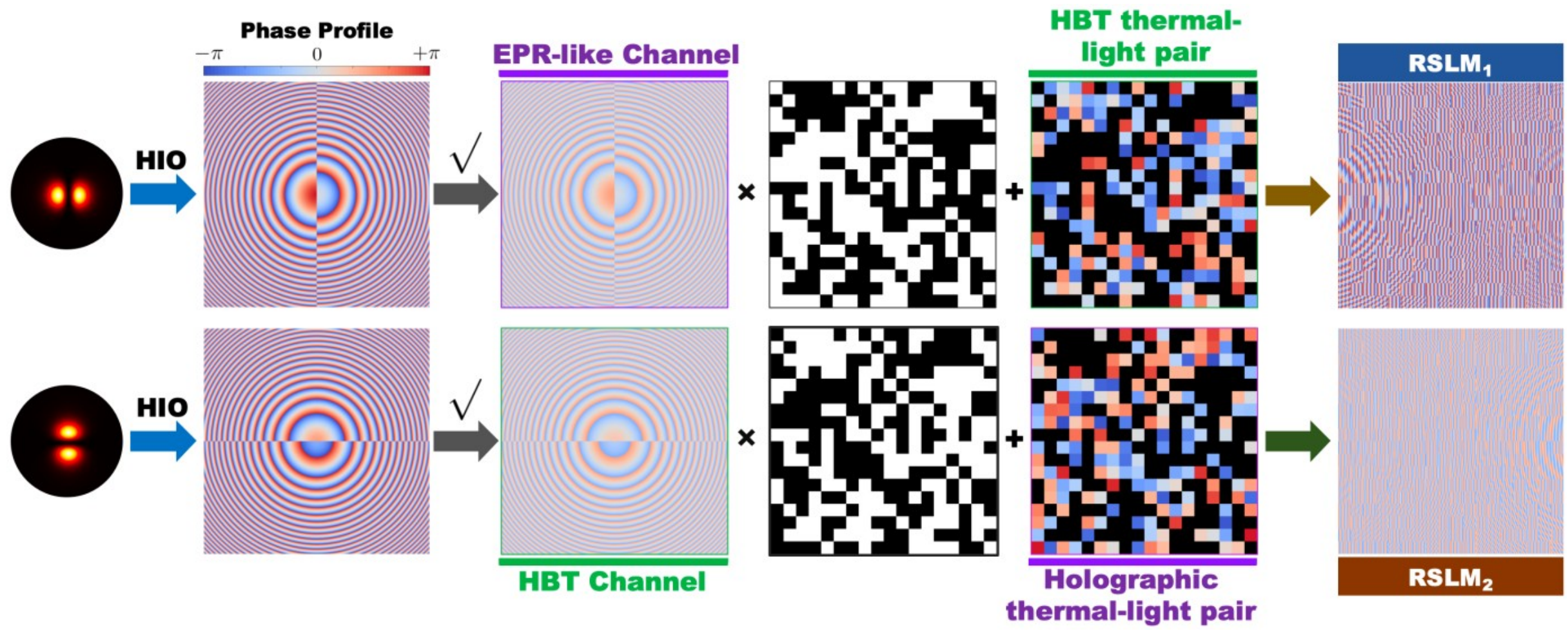

**Fig. 7. Image encoding method for two-photon spatial multiplexing.**

### Generation of the stray-light background

The experimental results presented in Fig. 3 are obtained from the encoding scheme in Fig. 7 and the corresponding calculated phases. The results shown in Fig. 4 are obtained from the optimized phases in Fig. S8. In addition, another He-Ne laser beam is used to illuminate $TSLM_2$ to generate a cat-shaped structured light beam, as shown in Fig. 5, which is then combined into one of the diffracted beams through an NPBS to serve as stray noise light. The design of the phase distribution loaded onto the $TSLM_2$ in the stray light beam is shown in Fig. S9.

## Acknowledgments

### Funding:

National Natural Science Foundation of China grant (12504411, Z. Y.) (12274037, J. X.); Basic and Applied Basic Research Foundation of Guangdong Province grant (2026A1515011704, Z. Y.); Science and Technology Development Fund of the Macao S.A.R. (FDCT) (grant 0105/2023/RIA2, H. C. L.) (grant 0002/2024/TFP, H. C. L.); Quantum Science Center of Guangdong-Hong Kong-Macau Great Bay Area (GDZX2204004, GDZX2304001, H. C. L.); Fundamental Research Funds for the Central Universities (Z. Y.).

**Author contributions:** Wanting Hou and Zhiyuan Ye contributed equally to this work. Conceptualization: Z. Y., J. X., K. W.; Writing-review and editing: Z. Y., J. X., K. W.;

Methodology: Z. Y., W. H., H. C. L.; Writing-original draft: Z. Y.; Data creation: W. H.; Investigation: W. H., Z. Y., H. C. L.; Validation: W. H., Z. Y.; Formal analysis: W. H., Z. Y., H. C. L., J. X., K. W.; Project administration: J. X.; Visualization: Z. Y., W. H.; Software: W. H., Z. Y.; Funding acquisition: Z. Y., J. X., H. C. L.; Supervision: J. X., K. W.

**Competing interests:** The authors declare no competing interests.

**Data, code, and materials availability:** The data that support the findings of this study are available from the corresponding authors upon reasonable request.